\documentclass[onecolumn,authoryear]{els-mrw} 

\usepackage{amsmath,amssymb,amsfonts,amsthm,makeidx,graphicx}
\usepackage{txfonts}
\usepackage{helvet,hyperref}
\usepackage{grantjournals}

\def\kpc{{\rm\,kpc}}
\def\eg{{e.g.,\ }}

\def\msun{{\rm\,M_\odot}}
\def\lsun{{\rm\,L_\odot}}

\newcommand{\simlt}{\lower.5ex\hbox{\ltsima}}

\newcommand{\simgt}{\lower.5ex\hbox{\gtsima}}

\begin{document}

\chapter{The satellite galaxies of the Milky Way and Andromeda}\label{chap1}

\author[1,2]{Amandine Doliva-Dolinsky}%
\author[2]{Michelle L. M. Collins}%
\author[3,4]{Nicolas F. Martin }%

\address[1]{\orgname{Dartmouth College}, \orgdiv{Department of Physics and Astronomy}, \orgaddress{Hanover, NH 03755, USA}}
\address[2]{\orgname{University of Surrey}, \orgdiv{School of Mathematics and Physics}, \orgaddress{Guildford, GU2 7XH, UK}}
\address[3]{\orgname{Universit\'e de Strasbourg, CNRS, Observatoire astronomique de Strasbourg}, \orgdiv{UMR 7550, F-67000 Strasbourg, France}}
\address[4]{\orgname{Max-Planck-Institut f\"{u}r Astronomie}, \orgdiv{K\"{o}nigstuhl 17, D-69117 Heidelberg, Germany}}
\articletag{Chapter Article tagline: update of previous edition,reprint.}

\maketitle

\begin{glossary}[Glossary]
\term{Baryonic matter}, in astrophysics,  refers to matter composed of protons, neutrons, and electrons, although electrons are not classified as baryons according to the particle physics definition.   \\
\term{Dark Matter} makes up 27\% of the mass-energy content of the Universe. Its composition remains unknown mainly because it does not interact via electromagnetic forces and is only detected by its gravitational effects on visible matter.  \\
\term{Satellite galaxy}, a galaxy that is gravitationally bound to a more massive galaxy.\\
\term{Dwarf galaxy}, a galaxy that has a stellar mass equal to, or lower than, $M_*=3\times10^9\,{\rm M_\odot}$.\\
\term{Local Group}, the group of galaxies within the local field that the Milky Way belongs to. It contains over 100 individual galaxies, and has a diameter of roughly 3 mega-parsecs (10 million light years).

\end{glossary}

\begin{glossary}[Nomenclature]
\begin{tabular}{@{}lp{34pc}@{}}
L$_\odot$ & The luminosity of the Sun\\
LG & The Local Group of galaxies \\
$M_*$ & The stellar mass of a galaxy\\
M$_\odot$ & The mass of the Sun\\
MW & The Milky Way galaxy.\\
M\,31 & The Andromeda galaxy.\\
$M_V$ &The absolute magnitude of a galaxy in the $V$-band.\\
$r_{\rm half}$ &The radius within which half the light of a galaxy is contained.\\
$\tau_q$ & The time at which star formation ceases within a galaxy\\
Parsec (pc) & Unit of distance (3.26 light-years)

\end{tabular}
\end{glossary}

\begin{abstract}[Abstract]
The satellite galaxies of the Local Group provide us with an important probe of galaxy formation, evolution, and cosmology. The two large spirals that dominate this group -- the Milky Way and Andromeda -- are each host to tens of satellites, ranging in stellar mass from $M_*=3\times10^9\,{\rm M_\odot}$ down to as little as $M_*\sim1000\,{\rm M_\odot}$. In this review, we (1) provide an overview of the known satellite population of the Milky Way and Andromeda, including how they are discovered and their observed properties; (2) discuss their importance in understanding the nature of dark matter, star formation in the early Universe, the assembly histories of their massive hosts, and the impact of reionisation on the lowest mass galaxies; and (3) highlight the coming revolution and challenges of this field as new observatories and facilities come online. In the coming decades, the study of Local Group satellites should allow us to place competitive constraints on both dark matter and galaxy evolution. 
\end{abstract}

\subsection*{Key Points and Learning Objectives}
\begin{itemize}
\item Overview the history of satellite galaxy detection in the Local Group.
\item Understand the defining characteristics of dwarf galaxies in the Local Group.
\item Obtain an overview of the structural and chemodynamical properties of dwarf galaxies.
\item Learn how dwarf galaxies are used as cosmic probes of dark matter and galaxy evolution.
\item Overview the future potential for finding dwarf galaxies and using them to test theories of dark matter.
\end{itemize}

\section{Introduction}\label{chap1:intro}

The Local Group (LG) provides us with a natural laboratory with which to study a diverse system of galaxies that trace a dynamic range of environments. It is dominated by two massive spirals -- the Milky Way (MW) and Andromeda (M\,31) -- each of which is host to dozens of low mass satellites \cite[\eg][]{mcconnachie08,mcconnachie18,martin09,martin13a,martin16,laevens15,drlica15,bechtol15, torrealba16b,cerny21a,cerny21b,cerny22,cerny23a,tan24}. These smaller companions range in stellar mass from $M_* \sim 3\times10^9 \,{\rm M_\odot}$, to as low as a few thousand solar masses. These low mass galaxies -- often referred to as dwarf galaxies -- are fundamental for our understanding of a range of physics including star formation, dark matter, and galactic and chemical evolution. The low mass satellites in the Local Group are close enough that we can study them star-by-star in both imaging and spectroscopy. This allows us to measure properties including their structural properties, mass profiles (both baryonic and dark), chemistries, star formation histories over a Hubble time and (in some cases) orbital properties in exquisite detail. Such a wealth of data makes the satellites of the Local Group an important population.

In this review, we will focus on galaxies that are considered to be satellites of either the MW or M31\footnote{Although we also include `satellites of satellites' \eg those galaxies that may have fallen in with the LMC}. At the time of writing, there are 88 confirmed satellite galaxies in the LG (49 in the MW and 39 in M\,31) and a further 15 candidate galaxies (14 in the MW and 1 in M\,31). These broad classifications can be separated further into satellites of satellites; the Large Magellanic Cloud (LMC) is thought to host 7 of the MW satellites \citep[\eg][]{fritz19,battaglia22,pace22} and Triangulum (M33) is the likely host of one M\,31 satellite galaxy (Andromeda XXII, \citealt{chapman13}) and one candidate satellite galaxy (Pegasus VII, \citealt{martinezdelgado22,collins24}). We define candidate discoveries as objects that cannot be classified as classical dwarfs or globular clusters based on the size-luminosity relation (see Section 2.1). Confirmation of the presence of a high dynamical mass from additional measurements (e.g. spectroscopic follow-up), indicative of a large to mass-to-light ratio and the presence of a dark matter halo, are essential to determine their nature. For some candidates, such as Ursa Major III/UNIONS 1 \citep{smith24}, spectroscopic follow-up remains inconclusive due to their low stellar counts and the velocity uncertainties, making it difficult to confirm the presence of a dark matter halo. We also note as candidates satellites of satellite, whose association with their host remains to be dynamically confirmed. In terms of morphology, they are composed of a range of galactic types, including the star-forming dwarf irregulars (dIrrs, LMC, SMC) and dwarf spiral (dS, M33), as well as the quenched compact and dwarf ellipiticals (cE, dE) and dwarf spheroidals (dSph). Additionally, the discoveries of very faint systems in the past two decades have led to an new galaxy ``type," that of ultra-faint dwarf galaxies that populate the faint end of the galaxy luminosity function. These are often defined as having $<10^5\,\textrm{L}_\mathrm{sun}$, as opposed to the brighter ``classical" dwarf spheroidals that populate the Local Group. 

The reader should, however, keep in mind that some of these classifications are more practical than physical and should not be construed as always defining dwarf galaxies with fundamentally different properties. For instance, the separation between dwarf ellipticals (originally found much earlier, outside the Milky Way surroundings and therefore intrinsically brighter) and the fainter dwarf spheroidal galaxies (with the same morphology but originally discovered as lower surface brightness Milky Way satellites that could not yet be discovered outside our immediate cosmic environs) is driven by the quality of the data used for their discovery. Similarly, "ultra-faint" dwarf galaxies owe much more to the advent of large CCD photometric surveys than to a fundamental physical difference with their brighter dwarf spheroidal counterpart. Indeed, these surveys enabled their discovery as over-densities of resolved stars whose surface brightness is below the sky brightness level, preventing their detection as unresolved "patches" of integrated lights in earlier surveys based on photographic plates. In what follows, we will favor the more generic term ``dwarf galaxies.''

\begin{figure*}
	\includegraphics[width=\textwidth]{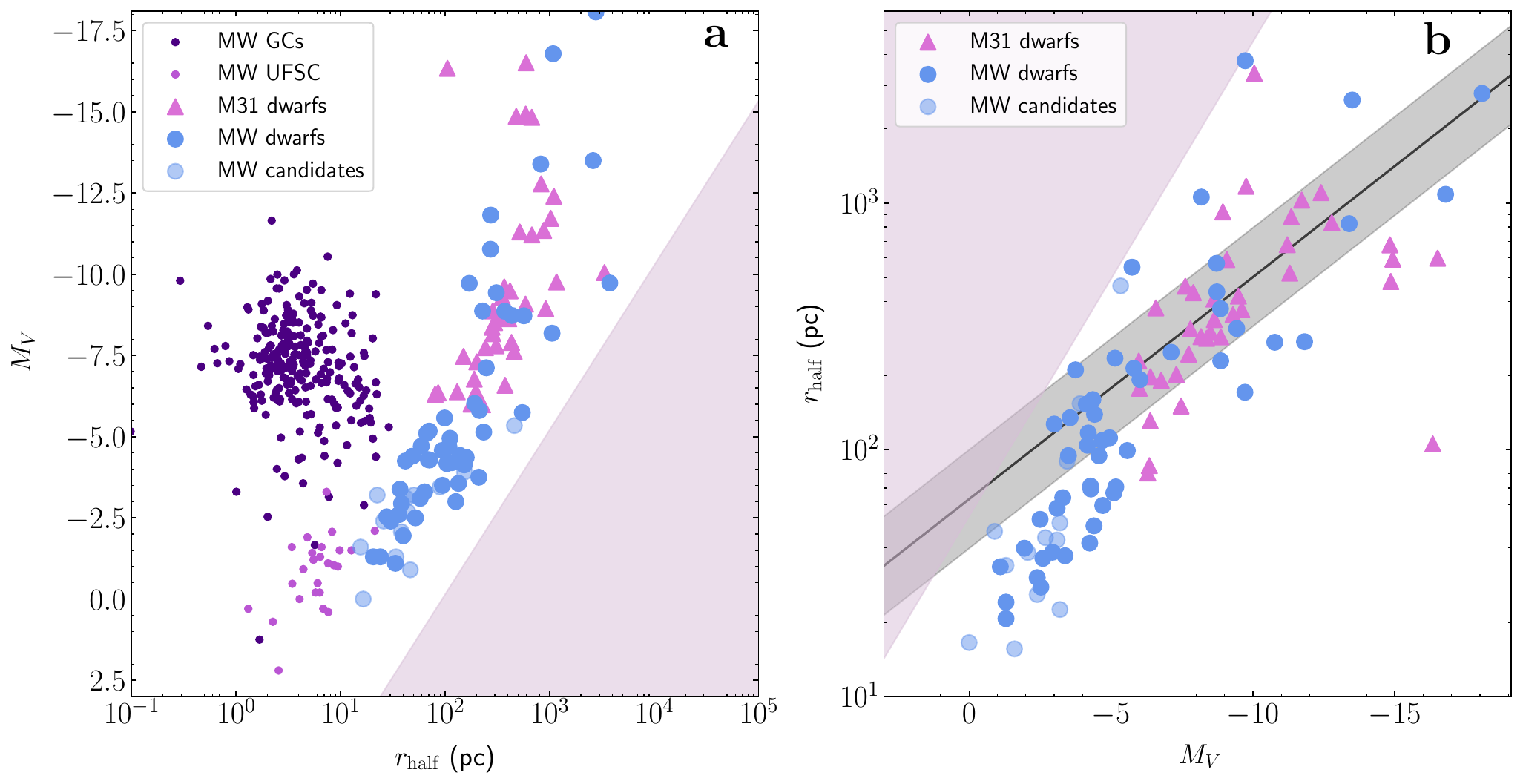}
      \caption{{\bf Left:} Half-light radius ($r_{\rm half}$) vs absolute magnitude $M_V$ for globular clusters (GCs, small purple points), MW dwarf galaxies (blue) and candidates (light blue) and M\,31 dwarf galaxies (pink). At bright magnitudes, there is a clear division between GCs and dwarfs, however below $M_V\sim-5$, this division is blurred. In this region, chemodynamical measurements are required to determine the nature of a stellar association. {\bf Right:} The size-luminosity relation for dwarf galaxies and candidates. The grey band shows a relation derived for all LG satellites from \citet{brasseur11} which well describes the bright end of this relation. Below $M_V\sim-6$, a departure or break is seen. This is believed to show the impact of reionisation on the lowest mass systems, which curtailed their star formation in the early Universe. In both plots, the shaded pink area shows an indicative surface brightness limit for current wide-field surveys in the Local Group.}
    \label{fig:morph}
\end{figure*}

In fig.~\ref{fig:morph} we show the size vs. brightness properties for all these systems alongside globular clusters (which overlap with dwarf galaxies at the low luminosity end). In this chapter, we aim to review the properties of the Local Group satellites, and how we can use them to interrogate physical and cosmological models. Two of the complexities of this field of research are directly visible in fig.~\ref{fig:morph}: (1) at the faint end, the clear size separation between the large dwarf galaxies and the compact globular clusters fades away (left panel; \citealt{gilmore07}), rendering the secure identification of dwarf galaxies significantly harder; and (2), also at the faint end, detection limits driven by the surface brightness of a satellite mean that a global view of dwarf galaxies is significantly more difficult to assemble as many satellites are likely below current detection limits \citep{drlica21,dolivadolinsky22}.

\begin{figure*}
	\includegraphics[width=\textwidth]{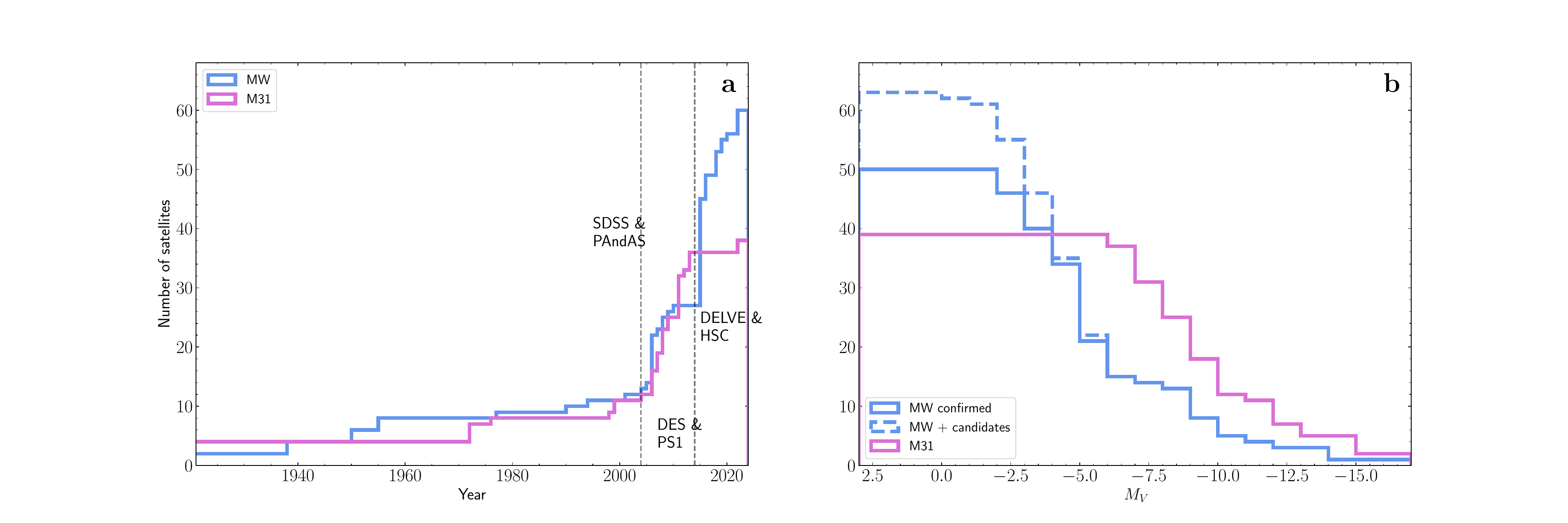}
      \caption{{\bf Left:} A timeline of the discovery of dwarf galaxies in the Milky Way (blue) and M\,31 (pink). The cumulative distribution is showns, and the impact of wide-field CCD surveys can clearly be seen after the year 2000, as many more dwarfs were discovered. {\bf Right:} The luminosity function of the Milky Way and M\,31 satellites.}
    \label{fig:timeline}
\end{figure*}

A running joke in the Local Group community is that to review the population of Local Group satellites is to undertake a fool's errand. Since the turn of the millennium, the population of known satellite candidates around both the Milky Way (MW) and Andromeda (M\,31) has more than quadrupled (for a timeline of this population explosion, see Fig.~\ref{fig:timeline}). With current and near-future surveys including but not limited to the DELVE, Euclid and the Vera Rubin Large Survey of Space and Time (LSST), this list will only increase further, with numerous works predicting 100s of new satellites will be discovered \cite[\eg][]{tollerud08,koposov08,walsh09,hargis14,kim18,newton18}. As such, it is very likely that, by the time this chapter goes to press, it will already be out-of-date.

The question then becomes `how to write a sensible review of the Local Group satellites that does not become immediately irrelevant?'. These authors endeavor to give you an overview of the current population, a discussion of the questions they pose about the nature of both dark matter and galaxy formation physics, and how we may use these small but mighty galaxies to better understand the Universe as a whole. And, to guard against immediate irrelevance, we also \href{https://github.com/michellelmc/Local-Group-Satellites/tree/main}{provide code for all figures made within this work}, and links to the various databases used to generate them. One particularly useful resource is the \href{https://github.com/apace7/local_volume_database}{Local Volume database}, constructed and maintained by Andrew Pace \citep{pace24}. This Local Bible should keep up-to-date with the onslaught of new detections and discoveries throughout the Local Group (LG) and beyond.

\subsection{A brief history of satellite galaxy detection}

The brightest satellites of the MW --  the Large (LMC) and Small (SMC) Magellanic Clouds -- are visible to the naked eye in the Southern Hemisphere, and have been an important part of human culture long before they were named after Magellan in the 1500s in Europe. They have luminosities of $3\times10^9\,{\rm L_\odot}$ and $9\times10^8\,{\rm L_\odot}$. As discussed in a review of the Clouds history by \cite{dennefeld20}, South American peoples have long referred to these objects as part of their oral history. The Tupi-Guaranis in Brazil referred to the clouds as fountains from which a tapir (LMC) and pig (SMC) are drinking; while the Mapuche of Chile describe them as ponds. The earliest recorded reference to the Clouds is believed to be from ``The Book of Fixed Stars'' by Abd al-Rahman al-Sufi in 964 CE. It would be almost a millennium before these two objects were first suggested to be separate satellites of the MW by \citet{abbe1867} who stated that \emph{``The visible universe is composed of systems, of which the Via Lactea, the two Nubeculce, and the Nebulae, are the individuals, and which are themselves composed of stars (either simple, multiple, or in clusters) and of gaseous bodies of both regular and irregular outlines''} (where Via Lactea refers to the Milky Way, and the two nubeculce are the LMC and SMC). There is currently a proposal to the International Astronomical Union to rename the Clouds, as they were not discovered by Magellan, who is widely known to have murdered and enslaved many Indigenous people in South America \citep{delosreyes23}. 

Perhaps remarkably, no further satellite galaxies were detected around the Milky Way until 1937, when the Sculptor dwarf was found by Harlow Shapley in 1937 \citep{shapley38}, using the 24-inch Bruce refractor at Boyden Observatory. Sculptor has an absolute magnitude of $M_V=-10.8$ (or $\sim2\times10^6\lsun$), meaning that it was the faintest galaxy yet detected by a considerable margin (compared with $9\times10^8\,{\rm L_\odot}$ for the SMC). He initially described it as a rich cluster of stars with ``remarkable properties''. Despite this being the first detection of such a galaxy, he prophetically noted that such systems may be quite common, as such low surface brightnesses would allow similar systems to remain undetected at that time. And indeed, over the next 60 years, more discoveries began to trickle in. Shapley himself discovered the similarly bright Fornax dwarf the following year \citep{shapley38b}. Then another 4 dwarfs (Leo I, Leo II, Draco and Ursa Minor) were detected between 1950 and 1970 \citep{harrington50,Wilson1955,cannon77}. Very little progress was made in the final few decades of the 1900s, with a handful of satellites being discovered through photographic plate imaging from the Palomar Sky Surveys and the ESO/SRC Southern sky surveys \cite[][]{irwin90,kk98}.

\subsubsection{The digital revolution -- expanding the family}\label{chap1:surveys}

Heading into the 21st Century, the Local Group satellite census stood at approximately two dozen galaxies \cite[\eg][]{mateo98}. However, over the past 2.5 decades, our view of the Universe has been revolutionised as changes in imaging technology have led to a wealth of discoveries of galaxies orders of magnitude fainter than what had been discovered previously  \cite[\eg][]{willman05a,willman05b,belokurov06a,belokurov07a,koposov15,laevens15, cerny21b,cerny22}, A key driver in this step-change was the adoption of CCD cameras in astronomical imaging. 

 The first charge coupled device (CCD) sensor was designed and invented in the 1969 by Willard S. Boyle and George E. Smith of Bell Labs. They went on to win the Nobel Prize in 2009 for their creation. The first astronomical images taken with a CCD were in 1976 with a 154 cm telescope on Mt. Lemmon \citep{smith76}. The advantages were immediately clear. Firstly, the image can be immediately viewed without complex processing. The quantum efficiency was far higher, allowing for longer exposures to detect lower surface brightness objects. And, as the data are recorded digitally, they can be immediately stored and read by any computer, something that cannot be said for large photographic plates. The first Voyager space telescope was launched in 1977 with CCD detectors on board, and returned incredible images of the outer planets \citep[\eg][]{smith79}.

CCD imaging was also implemented in wide-field ground based surveys such as the Sloan Digital Sky Server \cite[SDSS,][]{gunn98,york00}, which enabled a slew of detection of faint Milky Way companions \cite[\eg][]{willman05b,belokurov06a,zucker04}, including the unusually small and faint systems; Willman 1 and Segue 1 \citep{willman05a,belokurov07a}. With half-light radii of $r_{\rm half}\sim20-30\,{\rm pc}$ and stellar masses of order $M_*\sim10^3\,{\rm M_\odot}$, they fall into a range of parameter space that had previously been known as a zone of avoidance between globular star clusters and dSph galaxies, with no known stable objects possessing a half-light radius between 30-130~pc \cite[\eg][]{gilmore07}. With subsequent wide-field surveys including PanSTARRS 1 \cite[PS1][]{chambers16}, the Dark Energy Survey \cite[DES][]{bechtol15,koposov15,abbott18}, the Hyper SuprimeCam Survey \cite[HSC][]{aihara18}, and the DECam Local Volume Exploration survey \cite[DELVE][]{drlica21}, this gap has disappeared at low luminosity and the number of known satellites has increased four-fold. While, prior to the SDSS, the faintest known galaxy was Sextans with $M_V\sim-8.5$ \citep{irwin90}, we now have confirmed galaxies at $M_V=-2.0$ and candidates with $M_V=+2.2$ as shown in the luminosity function presented in fig.~\ref{fig:timeline}. If these turn out to be real galaxies, what does that mean for galaxy formation?

\subsubsection{Finding the satellites of M\,31}

The search for satellites in our nearest neighbour spiral has followed a largely similar path to that of the MW. The first recorded observation of a dwarf galaxy outside of the Milky Way (MW) was made in 1654 by Hodierna \citep{Hodierna1654} as he reported a cloud-like object, near the Triangulum constellation, that was later rediscovered by Messier in 1764 \citep{Messier1781}. This object is now known as the Triangulum galaxy or M\,33 and is considered to be the brightest satellite of M\,31. Later in the 18$^\mathrm{th}$ century, three other M\,31 satellites were also detected: NGC\,221, initially discovered by Guillaume Le Gentil in 1749, was later classified as M\,32 in Messier's catalogue in 1781; NGC\,205, first discovered by Caroline Herschel and then reported by her brother, William Herschel, in 1785 \citep{Herschel1785}\footnote{Messier did not initially classify this dwarf galaxy as it was present in his drawing of the Andromeda Nebula, but he later added it to the list as M\,101.}; and NGC\,185, discovered by Herschel in 1787. John Herschel, following in the family legacy, later discovered NGC\,147 in the 19$^\mathrm{th}$ century and Lewis Swift closed these initial discoveries of what would later be known as dwarf galaxies satellite of Andromeda with the detection of IC\,10.

With the brightest satellites of M\,31 catalogued, it then took almost a century for new discoveries to be made in the late 20$^\mathrm{th}$ century, thanks to the introduction of large surveys based on photographic plates. The Palomar Observatory Sky Survey (POSS) was instrumental in the discovery of fainter MW satellite dwarf galaxies \citep{Wilson1955}, which naturally led to the expectation that other M\,31 satellites remained to be discovered. But, the distance to these objects and their corresponding faintness meant that no such object was discovered on the survey plates. van den Bergh therefore conducted specific new observations using a set of nine highly sensitive plates and, indeed, they led to the detection of three new (candidate) satellites: And\,I, And\,II, and And\,III \citep{Vandenbergh1972}. These discoveries were complemented in 1976 by Karachentseva \citep{Karachentseva1976,Kowal1978} with the ``Local Group Suspect''\,3, or LGS\,3, later recognized as the Pisces dwarf galaxy, satellite of M\,31.

Following these discoveries, a span of 20 years elapsed before new dwarf galaxies were uncovered using the second version of the POSS II survey, in 1999. Based on the digitalized version of the survey, Armandroff employed a matched-filter technique to successfully identify two previously unknown dwarf galaxies, And\,V and And\,VI \citep{Armandroff1999}. In parallel, Karachentsev employed morphological criteria based on the POSS II plates to posit the existence of And\,VI and And\,VII \citep{Karachentsev1999}. These significant findings increased the total count of M\,31 dwarf galaxies to 11\footnote{The reader may wonder why And~IV and And~VIII are not listed as M\,31 dwarf galaxies despite their name. And~IV was initially thought to be an Andromeda satellite \citep{Vandenbergh1972} but later confirmed to be significant more distant \cite{Ferguson2000}. And~VIII was thought to be a new dwarf galaxy based on an overdensity of planetary nebulae \cite{Morrison2003} that was then shown not to be significant based on a wider mapping of these tracers in the M\,31 surroundings \cite{Merrett2006}.}, which, somewhat satisfactorily for two hosts with fairly similar properties, was comparable to the number of dwarf galaxies observed around the MW at that time. 

Over the last 20~years, our understanding of the M\,31 halo and dwarf galaxy system have been spurred by the combination of two main changes: (1) systematic and homogenous photometric surveys of large regions around M\,31 that reach significantly below the tip of the RGB at the distance of Andromeda ($\sim780\kpc$ \citealt{conn12}) and (2) the possibility to efficiently obtain large numbers of spectra for these RGB stars from multi-object spectrographs on 10-meter class telescope (in particular DEIMOS, on Keck~II \cite{Faber2003}).

Initial photometric mappings conducted systematically over the inner halo of M\,31 using CCD cameras \citep{Ibata2001,Ferguson2002} were expanded over time with wide panoptic surveys that probed the regions around M\,31 as part of MW surveys (e.g., SDSS, Pan-STARRS1) and led to the discovery of some new dwarf galaxies and stellar structures \citep{Zucker2004,Zucker2004b,Bell2011,Slater2011,martin13b,martin13a}. However, the most substantial advances in our understanding of the M\,31 dwarf galaxy system emerged from two complementary surveys: the Pan-Andromeda Archaeological Survey (PAndAS) and the Spectroscopic and Photometric Landscape of Andromeda Stellar Halo (SPLASH).

PAndAS is a systematic mapping of the region within $\sim150\kpc$ of M\,31 and $\sim50\kpc$ of M\,33 conducted with the MegaCam wide-field imager on the Canada-France-Hawaii Telescope \citep{mcconnachie18}. Its data reach $\sim3$~magnitudes below the tip of the RGB, which allows for the discovery of much lower surface-brightness features than was previously possible \citep{Ibata2007,Ibata2014}. The survey has led to the discovery of more than 15 new dwarf galaxies in the halo of M\,31 \citep{Martin2006,Ibata2007,mcconnachie08,martin09,richardson11}. Complementary spectroscopic data obtained with the DEIMOS spectrograph helped provide a more complete picture of these systems \citep{Chapman2006,collins13}. Following a different strategy, SPLASH focuses on the mapping of 50 fields scattered throughout the M\,31 halo, for which gravity sensitive photometry was obtained to more easily isolate M\,31 RGB stars from the foreground MW contamination. It led to the serendipitous discovery of a new satellite \citep{majewski07}. These fields were then preferentially targeted for DEIMOS spectroscopy and generated a sizable sample of bona fide dwarf galaxy giants that further enhanced their dynamical and metallicity portrait \citep{Gilbert2012,tollerud12,Gilbert2014,Gilbert2018}.

\subsection{Defining a galaxy -- star clusters vs dwarfs}\label{chap1:galdefine}

There are two immediate questions posed by the satellite galaxy luminosity functions shown in fig.~\ref{fig:timeline}. The first is `what is the faintest galaxy we can form?'. The faintest candidate known to-date -- Ursa Major III/UNIONS I -- has a (current) stellar mass of only $M_*=16^{+6}_{-5}\,{\rm M_\odot}$ \citep{smith24}. Can objects like this really qualify as galaxies? The second, more philosophical, question is: `what IS a galaxy?'. The surveys discussed above have allowed us to uncover a wealth of low mass, low surface brightness structures throughout the Galactic and M\,31 stellar halos, some of which are classified as galaxies while the others are classified as stellar cluster. Below a stellar mass of $M_*\sim 10^4\,{\rm M_\odot}$, we can no longer neatly divide these populations based on size and stellar mass alone. So how do we define a galaxy? And at what luminosity, if any, should we stop?

This seemingly innocuous question has divided the scientific community on multiple occasions over the past two decades. In the early 2010s, several sought to provide a community definition that would unite the field. Both \citet{forbes11} and \citet{willman12} attempted to establish a series of criteria required by a system to be considered a galaxy. The latter succinctly summarises that:

\begin{quote}
\large{
\quotehead{Galaxy, defined}
    ``A galaxy is a gravitationally bound collection of stars whose properties cannot be explained by a combination of baryons and Newton’s laws of gravity"
\source{\citet{willman12}}}
\end{quote}

\citealt{forbes11} additionally specifies that dwarf galaxies should have formed stars over a long enough period (of order Gyr) that a spread in iron abundance amongst it stars should be apparent. In essence, combining these definitions gives a chemo-dynamical definition where there are two key requirements to classify an observed group of stars as a galaxy rather than a star cluster:

\begin{enumerate}
    \item The stellar component is deeply embedded within a dark matter halo. These stars can be thought of as `collisionless', with their motions largely governed by the underlying dark matter distribution.
    \item Star formation (and hence chemical enrichment) should be extended over timescales of order Gyr, allowing a spread in iron amongst its stellar population. 
\end{enumerate}

In the current CDM paradigm, this translates to galaxies being collections of stars forming in the center of a dark matter halo. These two criteria whittle out almost all\footnote{There are a few globular star clusters that show a spread in their iron abundances, including $\omega$-Centauri, M54, Terzan 5 and Liller 1. These are considered to be the surviving remnants or nuclear star clusters of more massive galaxies, and as such aren't ``true'' globular cluster.} star clusters, where the motions of stars are governed purely by baryonic mass, are collisional and rarely show a spread in iron abundance within a single cluster (although see \eg \citealt{lee99,ferraro06,bellini09,carretta10,ferraro21} for a more detailed discussion of this). It also allows for physics beyond CDM, as modified gravitational theories are not excluded. Finally, it can be considered as a theory based on the evolution of a system rather than its present appearance. Star clusters form all their stars (almost) simultaneously and are bound by the gravity of their baryonic components. Galaxies, on the other hand, can form stars over an extended period of time, with their dark matter halos allowing them to hold onto and accrete gas over time, providing for multiple generations of stellar populations.

In general, this is a widely accepted definition, but it is still unsatisfactory to many. Most notably because it does not define a lower limit for a galaxy, nor does it explain what we would call a single star cluster that has formed within a dark matter halo. Such objects would be challenging to detect, but are now predicted to exist in simulations (\eg Taylor et al. submitted). However, this definition does allow us to separate our observed faint stellar systems into three categories:

\begin{enumerate}
    \item Dwarf galaxies - to be considered a dwarf galaxy, the stellar system must either have a dynamical mass measurement that cannot be explained by baryons alone (\eg a dynamical mass-to-light ratio $M/L>5\,{\rm M_\odot/L_\odot}$), or a spread in iron abundance greater than $\sigma_{\rm [Fe/H]}>0.1$~dex.
    \item Star clusters - systems where their dynamical mass agrees with their stellar mass, and that show no spread in iron abundance.
    \item Candidates - systems where observations do not allow us to rule out either a high mass-to-light ratio or a spread in iron.
\end{enumerate}

\noindent Many of the faintest detections fall into category (3) as they possess very few bright stars that can be followed up spectroscopically. This will continue to be a challenge even in the era of 30-m telescopes, and we will return to this later. In the remainder of this Chapter, we will focus only on the dwarf galaxies and candidates, leaving star clusters for another set of authors.

\section{Observed relations and trends}\label{chap1:obstrends}


Given their proximity to us, we can collect a vast amount of detailed information on the resolved stellar populations of Local Group dwarf galaxies and (where present) their interstellar medium (ISM). Within the Milky Way, we have access to the highest fidelity data, enabling us to measure the 3D positions, 3D velocities (radial and proper motions), masses and chemistries of individual stars. This allows not only for a detailed characterisation of their present day properties, but also their fully resolved star formation histories and chemical evolution \cite[\eg][for detailed reviews]{tolstoy09,mcconnachie12,fritz18,simon19,battaglia22,belokurov22}. Moving further afield, we lose some of these details. Notably it is much harder to obtain proper motions for systems beyond the edge of the Milky Way, although it is still possible with multi-epoch deep Hubble Space Telescope (HST) data \cite[\eg][]{sohn20,bennet23,casetti24}. 

Proper motions aside, we are able to construct a number of scaling relations for Local Group satellites from their sizes, luminosities, masses, chemistries and star formation histories. In this section we outline the detailed observations for these systems before moving on to their wider interpretation in section~\ref{chap1:probes}.

\subsection{Size vs Luminosity}\label{chap1:rhmv}

In fig.~\ref{fig:morph}, we show the absolute magnitude of Milky Way and M\,31 satellites vs. their measured (2D) half-light radii ($r_{\rm half}$). For all satellites there is a clear log-log trend between these parameters, with brighter galaxies being more extended. The first comparative analysis of the satellite systems of the MW and M\,31, conducted by Armandroff \citep{Armandroff1994}, revealed that their dSphs follow the same size-luminosity and luminosity-metallicity relations. Consequently, the author suggested that despite the different masses of the two hosts, they both offered a somewhat similar environment for the formation and evolution of satellites. However, as we have added more satellites from both systems to this relation, multiple authors have investigated the shape of this relation and its associated scatter, both jointly for the MW and M\,31 \cite[\eg][]{brasseur11}, and separately for both \cite[\eg][]{brasseur11,Manwadkar2022,dolivadolinsky23}. While the general trend is similar, we observe some differences for the brightest satellites ($M_V<-10$). On average, the M\,31 satellites appear slightly more extended than their MW counterparts; however, as M\,31 also has $\sim3\times$ the number of bright dSph satellites as the MW, it is difficult to interpret whether this is driven by some underlying physical or environmental difference, or just that they better probe the full scatter of this relation. 

An interesting feature of this relation is that at low luminosities ($M_V>-6$), there is a clear break. This is mostly traced by the MW satellites, as this is right at the current detection limit for satellites around M\,31 within the PAndAS footprint. However, the handful of satellites discovered with $M_V\sim-6$ in M\,31 also start to show a departure. In \citet{Manwadkar2022}, they find that this break occurs at the mass scale where dwarf galaxies become very sensitive to reionisation. These low luminosity systems have thus formed the bulk of their stars prior to reionisation, resulting in a steeper slope in the size-luminosity relation. This also agrees with the star formation histories of these systems, which show they are preferentially quenched early \cite[\eg][]{weisz14,brown14,savino23}. 

Another interesting feature is the population of clear outliers above this relation. In particular, there are three objects that deviate significantly: Crater II and Antlia II in the MW, and Andromeda XIX in M\,31. Prior to the detection of Andromeda XIX in \citet{mcconnachie08}, most spheroidal/elliptical dwarf satellites of the Local Group had scale radii of a few hundred parsecs. However, Andromeda XIX was initially declared to have a half-light radius of $r_{\rm half}=1.8$~kpc, almost a magnitude more extended than dwarfs of comparable luminosity ($M_V\sim-10$). This scale radius was subsequently remeasured to be $r_{\rm half}=3.1^{+0.9}_{-1.0}$~kpc in a homogeneous analysis of the structural properties of M\,31 dwarf galaxies \citep{martin16}. Crater II --- a dwarf galaxy with $M_V\sim-8$ and $r_{\rm half}=1.1\pm0.1\kpc$ --- was then discovered in the VST ATLAS survey a few years later \citep{torrealba16a}, and then Antlia II --- a direct analogue of Andromeda XIX --- was detected by \citet{torrealba19}. These three satellites were unlike anything previously seen, and were not predicted by any cosmological simulations of MW-like galaxies. Subsequently, they have also been shown to possess unusually low central masses and densities \cite[\eg][]{collins13,caldwell17,collins20}. It is suspected that these three ``diffuse giants'' have interacted substantially with their host galaxies in the past \cite[\eg][]{amorisco19,collins20,ji21}. As such, they are brilliant probes of the processes of tidal stripping and shocking on low mass systems. 

Similarly large, low surface brightness galaxies like Antlia II, Crater II and And XIX are now being discovered further afield in large numbers \cite[][]{koda15,mihos15,martinezdelgado16}, and their nature and formation is a hot topic \cite[\eg][]{chan18,grishin21,zemaitis23}. These nearby objects are thus extremely important for understanding the larger population given the exquisite data we can collect in the LG.

\subsection{Mass vs. Luminosity --- the dark matter connection}\label{chap1:dm}

Most of the satellite dwarf galaxies of the LG are dispersion supported, with negligible rotation. The exceptions are the more massive satellites (LMC, SMC and M33) who all have significant rotation. These bright satellites have well-measured mass profiles from their HI rotation curves \citep[\eg][]{alves00,corbelli14}. For the dSphs and dEs, their central masses can be derived using the velocity dispersions of their stars using the Virial theorem. A number of mass estimators exist to turn a velocity dispersion measurement, $\sigma_v$, into a measure of the enclosed mass at a given radius. These include:

\begin{enumerate}
    \item the \citet{walker09} estimator which measures the enclosed mass within $r_{\rm half}$:
    \begin{equation}
        M(r_{\rm half})=A\,r_{\rm half}\sigma_v,
        \end{equation}
        where $A=580 \,{\rm M_\odot\, (km^{-2}s^2\, pc^{-1}}$)
    \item  the \citet{wolf10} estimator which also measures the mass enclosed within $r_{\rm half}$:
    \begin{equation}
    M(r_{\rm half})=3\,G^{-1}\sigma_v^2\,r_{\rm half}.
    \end{equation}
    \item the \citet{errani18} estimator which measures the mass enclosed within $1.8\times r_{\rm half}$:  
    \begin{equation}
    M(\rm 1.8 \times r_{\rm half})=3.5\,G^{-1}\sigma_v^2\,1.8\times r_{\rm half}.
    \end{equation}
\end{enumerate}

\begin{figure*}
	\includegraphics[width=\textwidth]{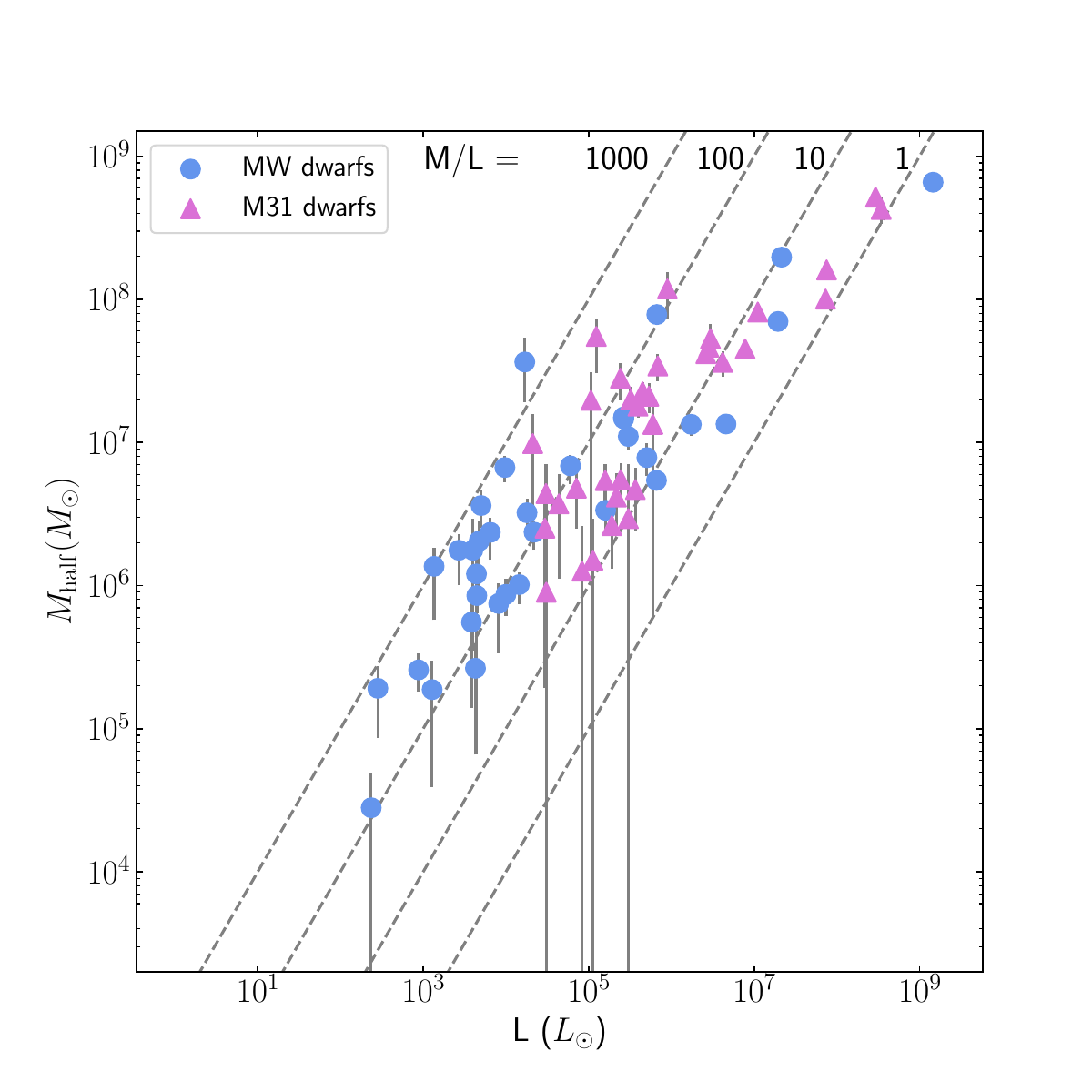}
      \caption{Mass contained within the half-light radius for dwarf galaxies is shown as a function of their luminosity. The dashed lines show mass-to-light ratios of 1, 10, 100 and 1000 ${\rm M_\odot/L_\odot}$ from right to left. The very brightest satellites are dominated by baryons in their centre, with $M/L\sim1-5\,{\rm M_\odot/L_\odot}$, while the fainter dSphs are seen to be much more dark matter dominated, wth $M/L>10\,{\rm M_\odot/L_\odot}$.}
    \label{fig:ml}
\end{figure*}

The first measurement of a mass-to-light ($M/L$) ratio for a dSph satellite was made from the velocities of just 3 carbon stars in the Draco system by Mark Aaronson  in 1983 \citep{aaronson83}. If these faint, old systems were formed only of baryonic material, they would be expected to posses $M/L\sim2\,{\rm M_\odot/L_\odot}$. Using the dispersion measured from these stars of $\sigma_v=6.5\,{\rm km\,s^{-1}}$, \citet{aaronson83} measured a  value of  $M/L=29\,{\rm M_\odot/L_\odot}$ for Draco, far higher than expected for a baryon dominated system. The implication was that these dSphs are dark matter dominated even in their very centres, far more so than their brighter galactic counterparts. In the decades since, the dSphs have been kinematically studied throughout the LG, allowing us to measure velocities for 10s-1000s of stars in these systems \citep[\eg][]{mateo91,mateo98,simon07,walker07,walker09,martin07,wolf10,tollerud12,tollerud14,collins13,collins20,collins21,charles23}. In fig.~\ref{fig:ml}, we see that the trend of enhanced $M/L$ in the dSphs remains true, and as you go to ever fainter luminosities, the dominance of dark matter increases. These results also suggest that, despite having luminosities that vary by a few orders of magnitude, the dwarf galaxies of the LG are mostly compatible with living in similar mass sub-halos. When we extrapolate these central measurements out to their virial radii ($R_{\rm Vir}$), they are consistent with halos of  mass $10^9$---$10^{10}\msun$. This has led to the discussion of a Universal halo for the faintest galaxies \cite[\eg][]{walker09,wolf10,strigari08}.

These high dark matter fractions have led to claims that the LG dSphs are ideal laboratories for testing different dark matter models. While all galaxies are expected to contain vast quantities of dark matter, the dSphs are unique in that they are not baryon dominated in their very centres. Even here, their large mass-to-light ratios persist. This means that their stars act as effectively massless tracers of the underlying dark matter potential. This is of cosmological interest as dark matter models make specific predictions about the structures of halos in their very centres. Early simulations of galaxies in the $\Lambda$-cold dark matter ($\Lambda$CDM) paradigm predict that they are extremely dense, with a strong central `cusp' \citep{navarro96a,navarro97}. Alternate theories, such as self-interacting dark matter (SIDM) predict lower central densities with flat `cored' profiles \cite[\eg][]{burkert00}. Warm dark matter (WDM) models are still predicted to be cuspy, but as structure forms slightly later in WDM, they are expected to be less concentrated than the CDM case \cite[\eg][]{colin00}. So, if we can measure detailed radial profiles for dark matter into the centres of these systems, meaningful constraints on the mass and self-interaction cross section of the DM particle may be feasible \cite[\eg][]{nadler24}.

Inevitably, this turns out to be more complicated than expected. Despite having few stars, the evolution of these faintest galaxies can modify the density profiles of their dark halos through stellar feedback and through their tidal interactions with their massive host galaxies \cite[\eg][]{navarro96b,read06, pontzen12,brooks14,read19b}. We will return to this in \S~\ref{chap1:cctbtf}.

\subsection{Luminosity - metallicity relation}\label{chap1:mvfeh}

There is a strong correlation between the luminosity (or stellar mass) of a galaxy and its average metallicity, which has been observed over many orders of magnitude \citep[\eg][]{mcclure68,tremonti04} including the low mass satellites of the LG \citep[\eg][]{mateo98,tolstoy09,kirby11,kirby13}. As you increase in luminosity (or stellar mass) you see a corresponding increase in the iron abundance, [Fe/H]. More massive systems have stronger gravitational potentials, meaning they are more able to retain metals and gas reservoirs throughout their evolution, resisting their expulsion from galactic winds and stellar feedback \citep[\eg][]{dekel86}. This means they are also able to form stars more efficiently and over longer timescales, further enriching their metallicities with time.  

\begin{figure*}
	\includegraphics[width=\textwidth]{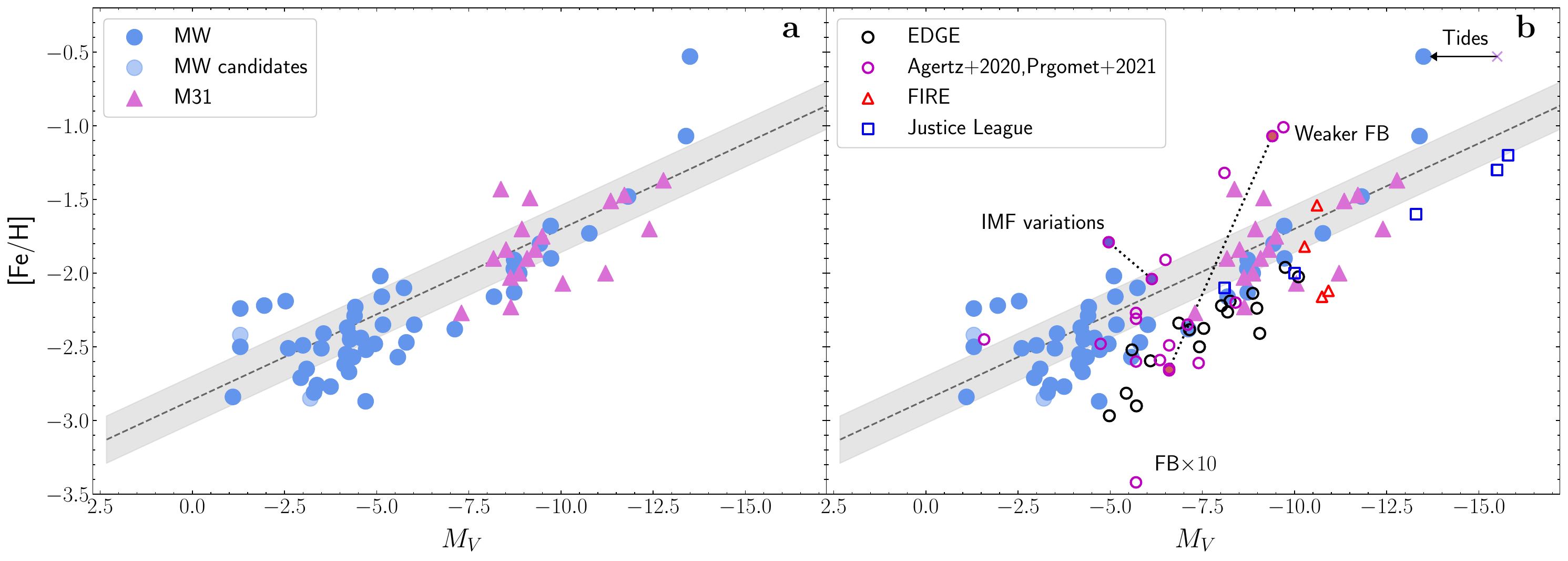}
      \caption{Both panels show the [Fe/H]-$M_V$ relation for dwarf galaxies of the LG. The black line and grey shaded band shows the fit and scatter in this relation as derived in \citet{kirby13}. At low stellar mass $M_V<-6$, we see a plateau, where metallicity no longer monotonically decreases. In the right panel, we add in simulations of dwarf galaxies and how well they match this relation. In general, simulations under-predict this relation. This relation is particularly sensitive to feedback models, with stronger/weaker feedback
moving galaxies vertically in this space (dotted line connecting the brown circles), and IMF variations moving galaxies
orthogonal to the relation (dotted line connecting the blue circles. Tides can also lower the $M_V$ of a system while its [Fe/H] remains unchaned, moving objects to the left in this diagram. Some or all of these processes may be responsible for the plateau seen at lower $M_V$.}
    \label{fig:feh}
\end{figure*}

We show this relation for the LG satellites in fig.~\ref{fig:feh}. The gray shaded region shows the best fit to this relation from \citet{kirby13}. This luminosity-metallicity relation is particularly informative at low masses as the metal enrichment of these small systems is very sensitive to stellar feedback \citep[\eg][]{jeon17,agertz20}. If stellar feedback is too efficient, gas and metals are easily removed from the galaxies, resulting in metallicities that are too low, while weak feedback results in metallicities that are too high. This can be seen in the right hand panel of fig.~\ref{fig:feh} where we show results from a number of different simulations including EDGE (\cite{agertz20,orkney21,prgomet22,kim24}); FIRE \citep{fitts17}; and Justice League \citep{applebaum21}. There are two points connected by a dashed line from the \citet{agertz20} simulations of a single galaxy run with two different feedback prescriptions, and one can easily see how stronger feedback results in dwarf galaxies that have metallicities that are too low, while weaker feedback results in galaxies that are too enriched. As such, this relation can be used to constrain feedback recipes in numerical simulations. In general, we see that most simulations under-predict the [Fe/H]-$M_V$ relation, which may imply their feedback recipes are too strong. 

Feedback is not the only physics that these systems are sensitive to. Both \citet{revaz18} and \citet{prgomet22} show that changes to the initial mass function (IMF) can also move a galaxy off the mass-metallicity relation and this may be particularly important at low masses. In \citet{geha13}, they find that the faintest dwarf galaxies in the MW ($M_V>-6$) show evidence for a bottom-light IMF, which may allow for a higher metallicity at a given mass. It is in the same luminosity regime that we start to see a departure from the mass-metallicity relation to a ``plateau'', where the faintest systems seem to be more metal rich than expected. However, this is also degenerate with the effects of tidal stripping, which would move a galaxy leftwards in this plot, i.e. lowering its luminosity while leaving its metallicity unchanged. Fully understanding these low mass galaxies -- particularly those within the plateau -- can therefore inform us about a wealth of galaxy formation physics.

\subsection{Star formation histories}

We are fortunate that all the satellites of the Local Group are close enough that Hubble (and now JWST) are able to resolve their stellar populations all the way down to their oldest main sequence turn-offs \cite[\eg][]{gallart05,tolstoy09,hidalgo11, deboer12,monachesi12,brown14,weisz14,monelli16,skillman17,savino23}. This depth allows us to piece together their precision star formation by comparing their observed colour-magnitude diagrams (CMD) with synthetically modelled equivalents, created using stellar evolution libraries \cite[\eg][]{tosi91,aparicio96,tolstoy96,dolphin97,gallart99,dolphin12}. These statistical methods allow us to infer the build up of these low mass systems, and understand whether and how they are affected by both internal and external factors, including star formation feedback, cosmology, reionisation and interactions with their massive hosts.

\begin{figure*}
	\includegraphics[width=\textwidth]{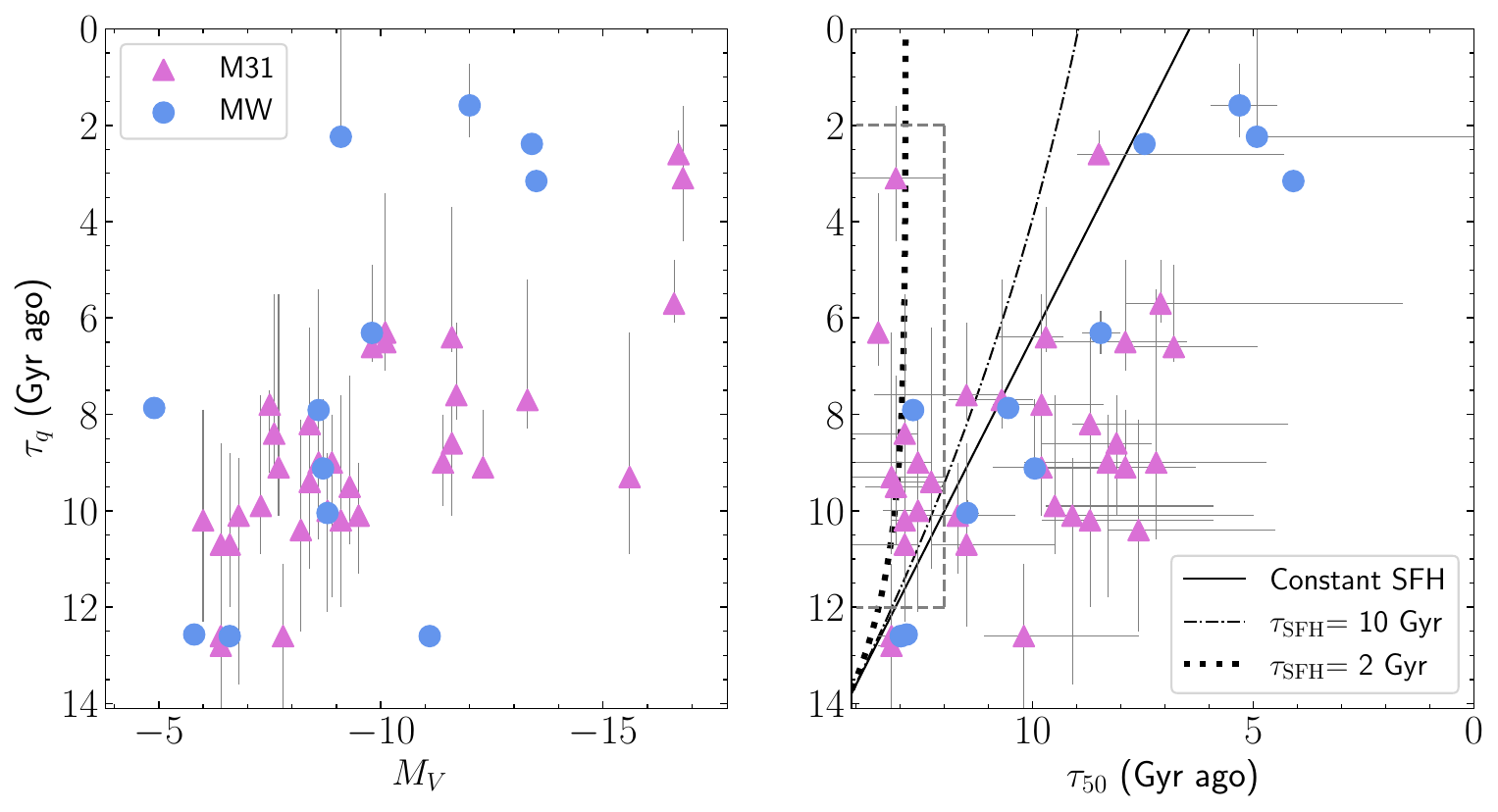}
      \caption{{\bf Left: } Absolute magnitude vs. quenching time ($\tau_q$, Gyr since star formation ceased) for MW and M\,31 dwarf galaxies. One can clearly see that as luminosity increases, the look-back time at which star formation ceases decreases, as massive galaxies are able to hold onto their gas reservoirs for longer than low mass systems. {\bf Right: } Time for 50\% of stars to form in the LG dwarf galaxies vs. $\tau_q$. The solid line shows how a galaxy with a constant SFR as a function of time would move through this plot, while the dot-dashed and dotted lines represent how galaxies with exponentially declining star formation over 10 or 2 Gyr would look. The dashed box highlights a region of parameter space where galaxies have old $\tau_{50}$ but are quenched at intermediate ages. M\,31 dominates in this region, with very few MW objects present.}
    \label{fig:sfh}
\end{figure*}

Other than the brightest (the MCs and M33), the satellites of the Local Group are no longer star-forming, and we refer to them as `quenched' systems \cite[\eg][]{brown14,weisz19,savino25}. They do show a range of quenching epochs that appears to be linked to their intrinsic luminosity, as discussed in \citet{weisz15} and Savino et al. 2024. In the left-hand panel of fig.~\ref{fig:sfh} we show the quenching times (defined as the time at which 90\% of the star formation has completed) vs. absolute magnitude to highlight this (data taken from \citealt{weisz15} and Savino et al. 2024). Interestingly, the M\,31 satellites also show a correlation between distance from their host and quenching time (Savino et al. 2024) while the MW satellites show no trend with distance \citep{weisz15}. 

The link between quenching time and luminosity is particularly fundamental. It has long been posited that, below a certain mass scale, reionisation can supress -- or completely quench -- star formation in dark matter halos. As galaxies started to switch on $\sim100$~million years after the Big Bang, intense UV radiation emitted by star formation caused the gas outside of galaxies in the Universe to transition from neutral to ionised. This process was largely complete $\sim1\pm$billion years after the Big Bang \cite[\eg][]{fan00,fan06,becker15,bosman22,becker24}. This effective heating of the Universe is expected to have negatively impacted the star-forming ability of the lowest mass dark matter halos \cite[\eg][]{efstathiou92,bullock00}. Reionisation essentially heats up the gas within these low mass galaxies, and as they have weak gravitational potentials, it is able to `boil' the gas out of them. This starves them of fuel for future star formation \cite[\eg][]{bullock00}. If they have not already formed (a significant amount of) stars, these halos will remain undetectable. However, there should be a transition population that formed enough stars to be detected by telescopes at the current epoch, but whose star forming properties show an abrupt transition at the epoch of reionisation. These are often referred to as ``reionsation fossils'' \cite[\eg][]{ricotti05,bovill09,bovill11a}.

A number of studies focusing on the faintest galaxies in the MW have shown that, indeed, there appears to be a common quenching period for galaxies with a stellar mass below $M_*\sim10^4\,{\rm M_\odot}$ \cite[\eg][]{brown14,weisz15}. As such, this is often referred to as a characteristic stellar mass scale for effective reionisation quenching. However, in M\,31 there are hints that things may be more complex than the MW population suggest. A significant number of M\,31 dwarfs show plateaus in their star formation histories around the epoch of reionisation, with star formation pausing for a few Gyr before then reigniting (Savino et al. 2024). This has not been observed in the MW system, and the most extreme case is that of the faint satellite, And XIII \citep{savino23}. This object has a stellar mass of $M_*\sim4.5\times10^4{\rm \, M_\odot}$, and would be expected to form its stars very early before being quenched by reionisation. However, quite remarkably, And XIII assembled only a small fraction of its stellar mass prior to reionisation, before experiencing a burst $\sim10$~Gyr ago in which 75\% of its stellar mass was formed. As such, this keystone object, and future detections like it, may give us an important insight into the interplay between reionisation and star formation in the lowest mass galaxies.

As we can derive detailed orbital histories for the MW satellites, we can look for trends in their star formation histories and their orbits. In \citet{miyoshi20}, they find that brighter dwarf galaxies show a peak in their star formation rate just after they are accreted, and that their star formation falls off more rapidly if they are on orbits which bring them closer to the MW. As we currently have proper motions for only a handful of M\,31 satellites \citep{sohn20}, we do not know if they also show this trend. 

\subsection{Comparing the MW and M\,31}
\label{sec:MW-M31_comp}

The sensitivity of dwarf galaxies to their environment and host merger history is widely acknowledged, though this has primarily been explored through simulations \citep{engler21,Kanehisa2023,Vannest2023} or through the detailed properties of individual MW or M\,31 dwarf galaxies \citep{weisz19}. In observations, only the impact on the brightest end of the satellite system has been studied \citep{Mao2021,Carlsten2022}, given the challenges of detecting faint satellites around more distant hosts. Nonetheless, \cite{dolivadolinsky23} inferred the properties of the M\,31 satellite system for dwarf galaxies with $M_V < -5.5$, thereby paving the way for a comparative study between the properties of M\,31's satellite system and that of the MW.

Firstly, it is worth mentioning that the comparison of both satellite systems highlights that dwarf galaxies in M\,31 stem from the same baryonic processes as those found in the MW. To first order, the global properties of the MW and M\,31 exhibit similarities. Specifically, the slope of the luminosity function falls within the same range. For the MW, $\beta$ = -1.9 $\pm$ 0.2 \citep{tollerud08} or $\beta \sim$ -1.25 \citep{koposov08}, while for M\,31, \cite{dolivadolinsky23}  infer a slope of $\beta$ = -1.5 $\pm $0.1. Furthermore, the results from \cite{dolivadolinsky23}  align with those of \cite{brasseur11} for the MW, illustrating that both systems of dwarf galaxies adhere to a similar size-luminosity relation.

Despite some statistical fluctuations due to the history of formation of a host, simulations show a consistent trend between the number of satellite dwarf galaxies and the mass of their host galaxy \citep{Starkenburg2013,engler21}. Consequently, if M\,31 is indeed more massive than the MW, one would expect that it should host more satellites than the MW if we assume that the two systems are representative of typical host galaxies. This prediction does align with the expected number of dwarf galaxies around both spirals of the Local Group. Model predictions for the MW estimate $\sim$20 to 40 satellites with $M_V < -5.5$ and $r_{\text{MW}} < 300$ kpc \citep{newton18, Jethwa2018, Nadler2020, Manwadkar2022, Drlica2020_1}. In contrast, M\,31 hosts 2--3 times more satellites: once dwarf-galaxy detection limits are taken into account \cite{dolivadolinsky22}, current observations constrain this number to $92^{+19}_{-26}$ satellite dwarf galaxies for Andromeda \cite{dolivadolinsky23}. In addition, M\,31 hosts a significantly higher number of globular clusters, with 92 globular clusters identified at projected distances between 25 and 150 kpc from M\,31 \citep{huxor14, Mackey2019}. This number is approximately seven times greater than the number of clusters found in a similar region for the MW. These tracers serve to reinforce the notion that M\,31 is indeed more massive than the MW.

A direct comparison with the $\Lambda$CDM simulation results implies a MW mass below $10^{12}$\,M$_\odot$ and an M\,31 mass $>1.5\times10^{12}$\,M$_\odot$. However, such a direct comparison is likely ill-advised as the exact number of (faint) satellite dwarf galaxies in simulations will depend strongly on the implemented baryonic physics and the accretion history of the host. Marginalizing over these effects in the latest generation of simulations may provide an additional constrain on the (relative) mass of the MW and M\,31.

Another compelling clue resides in the Star Formation Histories (SFHs) of the M\,31 and MW dwarf galaxies, as they present different patterns highlighted in Figure~\ref{fig:sfh}. M\,31's dwarf galaxies exhibit extended SFHs, with the majority of satellites quenched between 3 to 6 Gyr ago. On the other hand, the MW's satellites show quenching that predominantly occurred around 9 Gyr ago, with only a few more recently quenched systems (see right hand panel of fig.~\ref{fig:sfh}). These differences suggest a potential correlation between the satellites SFHs and the host merger history. One possible explanation for the surge of quenching of M\,31's dwarf galaxies between 3 to 6 Gyr ago is their environment, as simulations predict a major merger for M\,31 $\sim2-4$ Gyr ago \citep{hammer18, DSouza2018}. Indeed, these satellites may have been environmentally quenched or may have fell in M\,31 halo with the merger. This merger event could have played a significant role in shaping the SFHs of M\,31's satellites, and re-enforce the idea of an active M\,31 merger history.

\subsection{Satellites of satellites}

The $\Lambda$CDM model predicts that dwarf galaxies should also have satellite systems \citep{dooley17,santos2022}. In a hierarchical model of galaxy evolution, these satellites serve as building blocks for their hosts. This raises key questions: How much do mergers of large satellites contribute to the formation of these systems? How do they impact the host galaxies? And what observable properties of satellite systems are influenced by these mergers?

For the MW, it's worth noting that it is currently undergoing a merger with the LMC. Identifying which satellites are associated with the LMC \citep{battaglia22} is challenging due to its position within the MW's halo. However, mergers like this bring additional satellite galaxies, increasing the total number of satellites for the host \citep{Nadler2020,joshi23}.

Moreover, this increase in satellite galaxies affects their radial distribution. \citep{carlsten20} demonstrated that systems hosting an LMC-like dwarf galaxy show a slightly more concentrated radial distribution. The merger with a massive satellite like the LMC is also responsible for displacing the Milky Way by an estimated 15-25 kpc from the centre of its dark matter halo \citep{garavitocamargo21}.

Similarly to what is observed around the LMC, there was a natural expectation that faint dwarf galaxies would be discovered in the vicinity of M\,33 as satellite of this large companion. However, the sole confirmed satellite candidate dwarf galaxy found around M\,33 is And\,XXII \cite{Martin2006,chapman13}, with the possible very recent addition of Pisces~VII \cite{Collins2023}. This small number of M\,33 dwarf galaxy companions (which parallels a similar dearth of distant satellite globular clusters \citealt{Huxor2009}) aligns with a scenario that involves a past encounter between M\,31 and M\,33, as most of M\,33's satellites could then have been stripped and deposited in the M\,31 halo \citep{chapman13}. Nevertheless, the occurrence of such an interaction is subject to debate. Some evidence suggests that a previous close interaction took place, leaving a distinct signature in the star and gas distribution of M33 \citep{mcconnachie09,putman09,Semczuk2018}. But it has also been proposed that the kinematics of M\,33 and M\,31 support a scenario in which M\,33 and M\,31 have not yet strongly interacted with, at most, a single interaction $\sim6\,$Gyr ago at a minimum distance of 100\,kpc \citep{patel17,vandermarel19}. Nonetheless, recent studies \cite{patel18} demonstrate that considering the coverage and detection limits of the PAndAS survey, the limited number of M33 satellites is not inconsistent with expectations derived from $\Lambda$CDM.

\section{Satellite galaxies as cosmic probes}\label{chap1:probes}

Since the first mass-to-light measurements were made for the dSph galaxy population, they have been touted as precision probes of the dark matter model. As discussed in \S~\ref{chap1:dm}, these objects remain dark matter dominated at their very centres, making their stars essentially massless tracers of the dark matter density at all radii. Their number can also place constraints on competing dark matter models \cite[\eg][]{moore99,klypin99,wang14,vogelsberger16,kim21}. Typically, in CDM, we expect more low mass dark matter halos than in warmer models (with a particle mass below $\sim10$~keV) or those where the particle experiences self-interaction over some cross section. As such, if we can make a complete census of the lowest mass satellites, strong constraints can be placed on the CDM model \cite[\eg][]{kim18}. If we can cleanly isolate signatures of dark matter physics from galaxy evolution physics, we can thus use the abundance and internal dynamics of dwarf galaxies to understand the nature of the dark matter particle. This has turned out to be trickier than expected, and has spawned a number of ``small-scale crises'' \cite[\eg][]{bullock17,simon19,sales22}, along with a plethora of methods by which to solve them. In this section, we will discuss some of these challenges and how dwarf galaxies may yet shine a light onto the dark sector.

\subsection{The Missing Satellite (non-)problem }\label{chap1:msp}

As noted by almost everyone within the field, this ``problem'' is not now --- nor has it ever really been --- a problem. The term ``missing satellite problem'' dates back to some of the earliest simulations of Milky Way like galaxies and their dark matter subhalos in the CDM paradigm. At that time, there were only 11 known MW, and 8 M\,31 dwarf satellites. But in the late 90s, both \citet{moore99} and \citet{klypin99} ran dark matter simulations that predicted that MW-mass galaxies should have halos teeming with dark substructures. In particular, \citet{moore99} found that the number of substructures with halos similar to the detected MW objects should be 50 times higher than observed. Similarly, \citet{klypin99} found that, above a circular veolicty of $\sim20\,{\rm kms}^{-1}$, there should be approximately 50 compared with the 11 observed. In either event, a clear discrepancy between the number of dark subhalos and luminous satellites existed.

However, the important distinction here is indeed in \emph{dark} vs \emph{luminous} substructure. In simulations, we can track everything. In observations, we can only count what we can detect. And this nuance was not lost even in these two works which coined the missing satellite problem: both Moore and Klypin are clear that one obvious cause for this dearth of satellites is that their baryonic content was too low to be detected in current surveys. They also appeal to the physics missing from their simulations for further explanation of these differences. Dark matter simulations do not contain any baryonic physics. There is no star formation, no feedback and heating from massive stars and supernovae, no gas and no information on the mass scales at which luminous galaxy formation may become inefficient. And, indeed, it is precisely the mix of observational detection and baryonic physics that can induce a discrepancy of orders of magnitude; and remove it entirely \cite[\eg][]{sawala16,kim18,read19b}.

As discussed in \S~\ref{chap1:surveys}, the number of known faint galaxies around the MW and M\,31 have more than quadrupled since the original missing satellite publications. And our understanding of the \emph{completeness} of such surveys has also dramatically increased \cite[\eg][]{koposov08,tollerud08,Drlica2020_1,dolivadolinsky22}. The faint satellites of the MW and M\,31 are typically detected through their resolved stellar populations, where survey teams apply matched-filter algorithms (or similar) to catalogue data in order to detect over-densities of stars that align in colour-magnitude space with old stellar populations \eg a red giant branch population, or main-sequence turn-off \cite[\eg][]{koposov08,drlica15,bechtol15,martin16}. The fainter your system, the fewer the stars, and in particular, the fewer the number of \emph{bright} stars. A galaxy like Segue 1 only has a few giant stars in its entire virial radius. It was only detectable in SDSS as it was nearby enough (distance of 23 kpc, \citealt{belokurov07a}) for the well-populated main sequence turn-off to be visible. If there was a Segue 1 at 50 kpc, SDSS would not have been able to detect it. Even with our deepest current surveys, finding a Segue 1-like object at the outer extremes of the MW halo is unfeasible. However, by combining the known luminosity function with the survey depths, coverage and completeness functions, one can predict how many Segue 1's we expect to find out to the virial radius of the MW with future surveys \cite[\eg][]{tollerud08,walsh09,kim18}. For Andromeda, we can use the luminosity function of dwarf galaxies found within the PAndAS survey to estimate how many satellites it has down to a limiting magnitude of $M_V\sim-6$ \citep{dolivadolinsky22,dolivadolinsky23}. To-date, these calculations suggest we should eventually find hundreds of luminous dwarf satellites around both hosts.

Further, as simulations have advanced to include the baryonic processes discussed above, we also find their predictions for the number of satellites come down to meet the data \cite[\eg][]{sawala16,brooks14,read19a,font21}. We now see that these are in perfect agreement down to $M_V\sim-6$. Below this, discrepancies begin to appear, but survey completeness continues to drive this. In the era of future surveys like the Vera Rubin Large Survey of Space and Time (LSST), we should be able to identify many of these faint systems. In the event we do not, it may be that we are ruling out CDM in favour of warmer or more self-interacting particles. 

In summary, \emph{there is no missing satellite problem}. There remain satellites we are yet to find, and a lingering question about how many of the very faintest galaxies we expect to see throughout the Local Group. This merely means that there is a broad spectrum for new discoveries in the coming decade.

\subsection{The central density problem(s)}\label{chap1:cctbtf}

The CDM model has long made predictions for the mass distribution of dark matter in the very centre of galactic halos \citep{navarro96a,navarro97}. In particular, these early works found that as CDM halos assemble through dissipationless hierarchical clustering, they possess a density profile whose shape does not vary with mass or initial conditions. This lead to the definition of the Universal Navarro, Frenk and White (NFW) profile which has a divergent cusp ($\rho\sim r^{-1}$) at its centre. However, this cuspy prediction is at odds with the mass profiles of low surface brightness galaxies (\eg \citealt{flores94} and, indeed, noted in \citealt{navarro97}). These systems are more compatible with a flat central density, or constant density core. As more galaxies at ever lower masses have been studied, this `problem' has persisted, and is referred to as the `cusp-core' problem \cite[\eg][]{colin02,spekkens05,ogiya11}. This can lead to an order of magnitude difference between the predicted central density and the measured one. A long-suggested solution to this particular problem is via stellar feedback, particularly from supernovae \cite[\eg][]{flores94,navarro96b}, however for a long time, it was debated whether the dSphs of the LG had enough stars to inject the required energy to modify a cusp to a core.

Several key papers revisited and addressed this in the early 2000s. While much of the early work had considered a single, violent event to transform a cusp into a core, newer papers began to consider repeated bursts and their impact \cite[\eg][]{read05,mashchenko06}. In particular, \citet{pontzen12} demonstrated how repeated energetic bursts of star formation and feedback create fluctuations in the central potential, even when only a few percent of the baryons are forming stars. The contraction of gas in the centre of a halo to form stars is followed by the blow-out of gas via supernovae, transferring energy into the collisionless dark matter particles. This causes a change in their orbit, gradually flattening a high density cusp into a core. The work of \citet{pontzen12} showed that this is irreversible and that it is possible even on the low mass scales of dwarf galaxies. The key requirement is the repeated bursts of star formation and feedback in these systems. 

This effect is now widely seen in hydrodynamical simulations of dwarf galaxies \cite[\eg][]{brooks14,read16,elbadry16,read19b,dutton19,koudmani24}. This finding is also neat as it provides predictions: only low mass galaxies with extended star formation (star formation time --  $t_{\rm SF}$  -- exceeding $\sim8$ Gyr) should be able to form a core in this manner. Thus, galaxies whose star formation is stopped (quenched) at early times should host pristine cusps. Work by \citet{read19b} demonstrated that for isolated galaxies (whose dark matter profiles should be unaffected by external tidal forces), those with extended star formation are indeed more cored, and those that quenched early are more cusped. 

More recent work analysing satellites of the MW and M\,31 (that are not tidally isolated) have shown that star formation feedback cannot be the whole solution. In particular, for Crater II \citep{torrealba16a} and  Antlia II \citep{torrealba19} in the MW; and Andromeda XXI \citep{collins21} and XXV \citep{charles23} in M\,31. These 4 dwarfs are all formed of primarily ancient stars, with little-to-no star formation in the past 6-8 Gyr \cite[\eg][]{weisz19}, and so the assumption is they should retain cusps. However, all of them are extremely low density in their centres, implying a more cored distribution. While this could be taken as a sign that the cusp-core problem persits, we instead argue that these intriguing outliers serve as excellent laboratories for studying other proposed core-forming mechanisms. The most likely candidate here is tidal forces. In particular, Carter II and Antlia II are on orbits that bring them very close to the MW centre \cite[\eg][]{fritz18,torrealba19,ji21}, and this could explain their extreme central densities \cite[\eg][]{read06,penarrubia08a,penarrubia09,errani17,errani22,errani23}. 

Another issue with the distribution of dark matter in the centre of low mass dwarf galaxies was highlighted by \citet{bk11}. When analysing satellite subhalos around MW-mass halos in dark matter only simulations, they found that the most massive were too dense to host the brightest satellites observed around the MW. In essence, it suggested that there should be multiple massive dark halos around the MW that are too big to have failed to form to stars, and yet are not observed. As such, this was named the `Too Big to Fail' problem (TBTF). The original paper posits that there are several potential solutions to this issue. One can appeal again to the baryonic physics that are not implemented in these simulations, particularly feedback from supernovae and massive stars. \citet{read16} demonstrate that TBTF can be entirely alleviated with supernova feedback, while \citet{brooks14} show that tides can also lower the central masses of satellites. Other proposed solutions could be that the MW mass is lower than we expect, as the mass of the most massive subhalo scales with host mass (although then the problem merely shifts to M\,31, as LG mass constraints won't allow both to be low mass). Or, it could hint at a significant issue with CDM and again suggest a shift to a warmer or self-interacting particle which naturally form lower density subhalos. In general, it is now believed that (for satellite galaxies at least) TBTF can be entirely reconciled with $\Lambda$CDM when baryonic physics are included \cite[\eg][]{sales22}, through combinations of the phenomena discussed above.

\subsection{The asymmetric distribution of Satellites}\label{chap1:planes}

\begin{figure}[t]
\includegraphics[width=.5\textwidth]{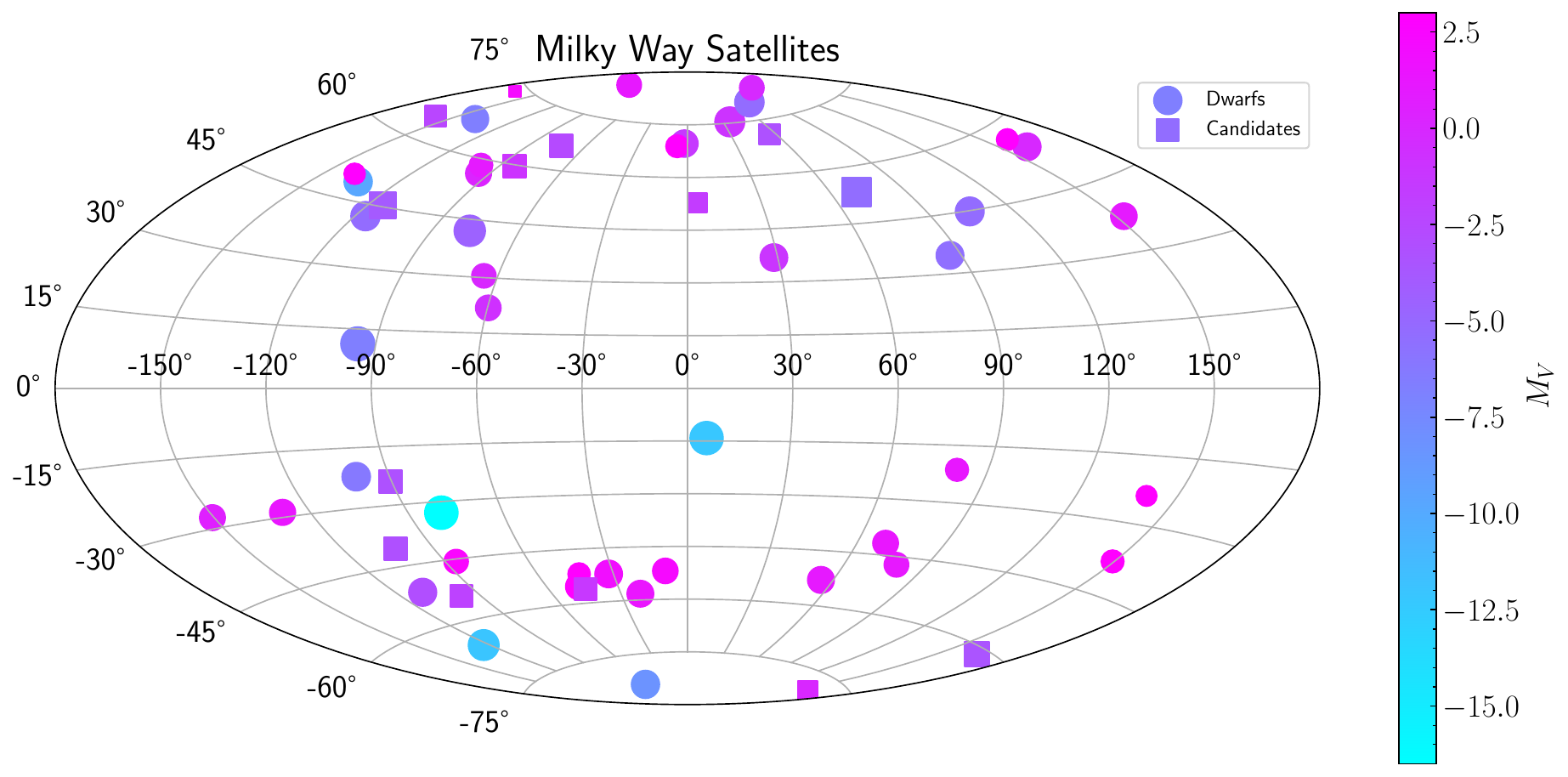}
\includegraphics[width=.5\textwidth]{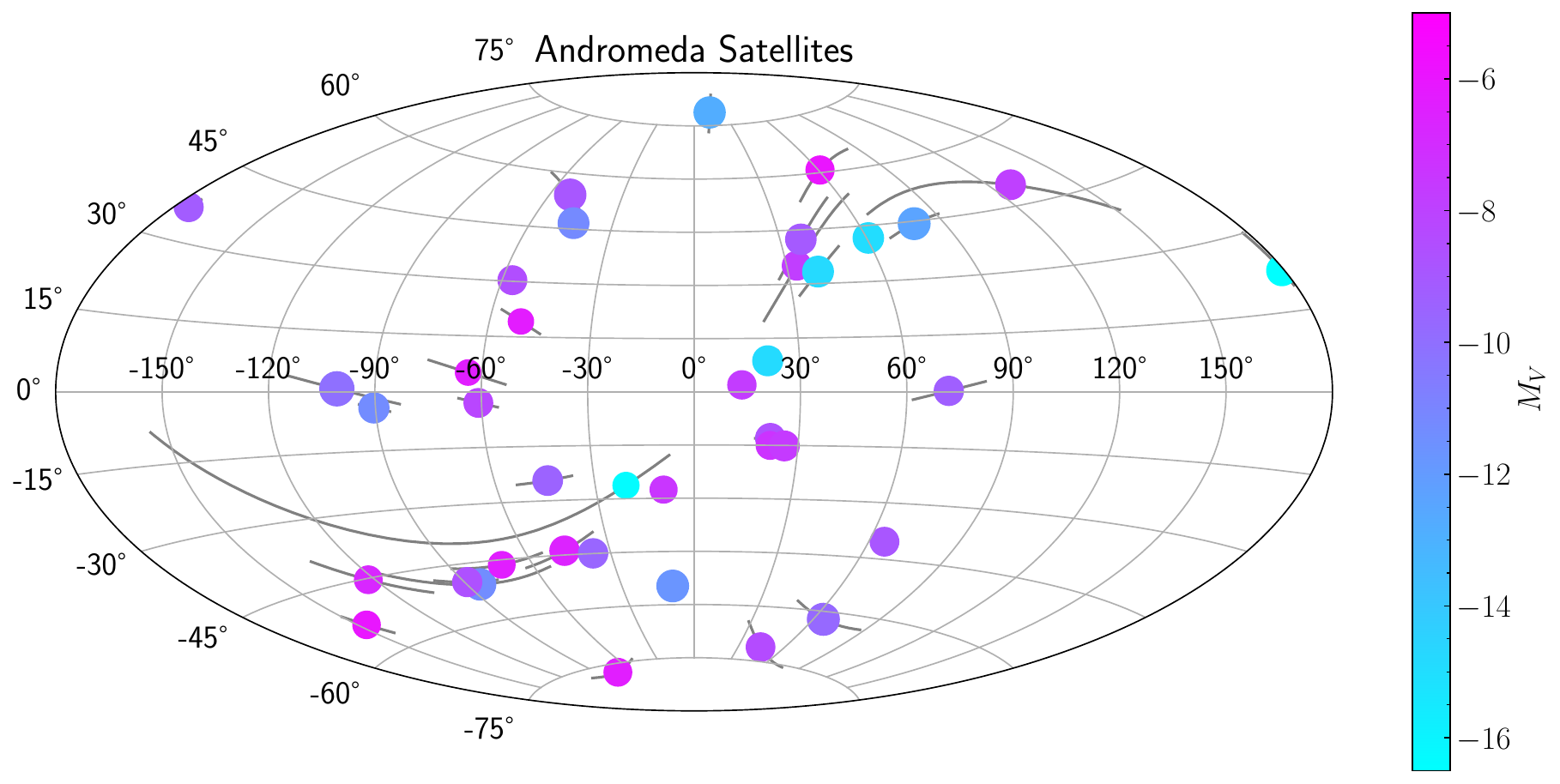}
\caption{Aitoff projections of the MW (left) and M\,31 (right) satellites around their hosts. They are colour-coded by their absolute magnitude. Neither distribution is isotropic, and both show planar distributions for $\sim50\%$ of their satellites. In both cases, these planes appear to be rotating, suggesting they are coherent structures rather than transient alignments.}
\label{fig:aitoff}
\end{figure}

Early simulations of MW galaxies and their subhalo populations suggested that satellites should be distributed isotropically around their hosts \cite[\eg][]{moore99,klypin99,springel08}. However, around both the MW and M\,31, it has long been noted that their satellite populations appear more anisotropic, see figure~\ref{fig:aitoff}. Claims that the MW satellites lie along a thin plane date back to \citet{Lynden1976}, while evidence that the M\,31 satellites are also non-uniform in their distribution date back to \citet{McConnachie2005b}. For the MW it is now widely recognized that a substantial portion of its dwarf galaxies exhibit a thin planar co-rotating distribution \citep[][]{Lynden1976, Kroupa2005, Metz2008,pawlowski12} even if the challenging nature of this feature to the $\Lambda$CDM framework is still debated \citep{Pawlowski2021,Sales2023,Sawala2023,Xu2023}.

The harvest of M\,31 satellites uncovered by the PAndAS survey \citep{mcconnachie09,mcconnachie18,martin16} showed without ambiguity that the M\,31 satellite system is also anisotropic. Similarly to what is observed around the MW, about half of the known dwarf galaxies of M\,31 distribute themselves in a plane-like distribution \cite[][]{ibata13,Pawlowski2013}. Among this sample of satellites, 13 of them exhibit co-rotation. An additional planar structures was also proposed \cite{Santos2020}, encompassing 18 satellites. Combined, these features include most of the M\,31 dwarf galaxies. Analyzing the transient aspect of these distributions proves challenging due to the lack of proper motion measurements for most satellites. Nonetheless, the few satellites with proper motions suggest that the system may exhibit dynamic coherence, as indicated by their determination of the proper motion of NGC 147 and NGC 185 \cite{sohn20}. In any case, the presence of these planes of satellites in Andromeda reinforces the notion that the distribution of dwarf galaxies around their host may not be isotropic.

The strong asymmetry in the distribution of M\,31 companions, with most of them located on the MW side of M\,31, is another remarkable aspect of the distribution of Andromeda satellite system that is not observed for the MW \cite{McConnachie2005b}. This distinct feature was further confirmed with increasingly refined distances based on the tip of the RGB \cite{conn12} and RR Lyrae stars \cite{savino22} and whether M\,31 is truly at the centre of its satellite system. Such a positional offset may stem from a similar phenomenon to what is observed in the MW, with the LMC pulling our Galaxy away from the centre of its halo \cite[\eg][]{erkal21}. In such a scenario, the fairly recent accretion of a giant satellite onto M\,31, producing the Giant Stream, would be an evident culprit for this observed distance offset.

 A plausible explanation for the plane-like distribution of dwarf galaxies around the MW or M\,31 \citep{Lynden1976,Kroupa2005,ibata13,Pawlowski2013} is the hierarchical accretion activity of their host galaxies. However, according to \cite{Kanehisa2023}, major mergers can enhance the phase-space correlation of satellites, with an observable signature lasting up to $2-5$ Gyr, but they may not create coherent and/or highly flattened planes of satellites. Even though the proper motion of M\,31's dwarf galaxies is currently unknown, this result suggests that the plane-like distribution of satellites around the MW and M\,31 may not be a signature of their merger history. However, the distribution of M\,31 dwarf galaxies display a strong anisotropy, as nearly all of them are between the MW and M\,31. This peculiar distribution may imply that the M\,31 satellites are not fully virialized, and the distribution could result from the sloshing of M\,31 due to a past merger event.

\section{The coming revolution}

As we described earlier, the release of photometric data from the SDSS in the early 2000s, followed by similar efforts to map the halos of both the Milky Way and Andromeda (DES, DELVE, PAndAS, Pan-STARRS1, \dots), was the start of a transformative era for our understanding of the Local Group dwarf galaxy system. Dozens of new dwarf galaxies were discovered over the past two decades, revealing satellites that vastly expanded our knowledge of what a dwarf galaxy can even be. These new discoveries have allowed for much more accurate comparisons between observations and predictions from galaxy formation in specific cosmological models. These more stringent constraints are stemming from the simple increase in the size of the samples of dwarf galaxies, the extended range of luminosities they now cover (including significantly more dark-matter dominated dwarf galaxies that are much more sensitive to both the properties of the dark matter particle and the details of baryonic physics), and the detailed follow-up photometry and spectroscopy that provide a much better understanding of each individual system.

Another such transformative epoch is upon us with the recent or approaching start of a new generation of photometric surveys that are expected to discover not dozens but a few \emph{hundred} next dwarf galaxy candidates. ESA's satellite Euclid has started mapping the Milky Way halo from space in February 2024. While Euclid is foremost a cosmology mission and performs photometric observing in a small set of large and visible/near-infrared filters that are not optimized for the study of old metal-poor stellar populations, it will nevertheless produce deep catalogues of stars over almost half the sky \citep{scaramella22}. These will undoubtedly lead to the discover of new Local Group dwarf galaxies.

Most prominent among the future projects, the Large Synoptic Sky Survey (LSST) will start observing the night sky in 2025 with the Vera Rubin telescope \citep{LSST09}. With its large 6.5-meter mirror and very large field of view, this telescope will provide very deep, stable photometry over half the sky to an unprecedented depth. The unrivaled compilation of the luminosity, the colors, and the motion of billions of stars will be a treasure trove for the discovery of new dwarf galaxies out to the edge of the Local Group \citep[and beyond,][]{mutlu-pakdil2021}. These new observations do come with their own sets of challenges, though. Beyond the difficulties of handling the massive volume of data that will need to be processed to discover small overdensities of a few tens or hundreds of stars with the right properties, the main complexity will emerge from the very challenge of separating stars from the much more numerous compact galaxies at faint magnitudes (beyond $\sim24^\mathrm{th}$ magnitude). Until now, this complexity was overcome by follow-up observations on larger telescopes, from deeper photometry or dedicated multi-object spectroscopy. In the era of the 6.5-meter LSST, few facilities will be available for such programs, especially given the expected discovery of hundreds of potential dwarf galaxy candidates that would require such follow-up.

New strategies are therefore required to confirm future dwarf galaxy candidates without the necessary requirement of follow-up, or to more specifically focus follow-up on targets that appear most valuable and interesting. One can think of combining the detection of an overdensity of star-like objects with the coincident detection of brighter tracers that are usually present in typical dwarf galaxies. Among those, (blue) horizontal branch stars and RR Lyrae are systematically present in all but the faintest dwarf galaxies \citep{baker15,martinez-vazquez19}. Another promising avenue is to leverage the luminosity-metallicity relation of dwarf galaxies and, since the most challenging dwarf galaxies to characterize would naturally be the faintest ones, to independently search for the coincidental presence of very metal-poor stars (\eg $[\textrm{Fe}/\textrm{H}]<-2.0$). Space surveys, such as Euclid mentioned above, or the future Chinese Space Station Telescope \citep[CSST; ][]{qu23}, with their much better resolution will enable a significantly better star-galaxy separation that is very complementary to LSST. Difficulties will remain for the faintest of system, like UNIONS~1/Ursa Major~III \citep{smith24}, especially at large distances, as they do not host any star brighter than their turnoff and, consequently, host no very metal-poor giant, horizontal, or RR Lyrae stars. The importance of such systems to constrain galaxy formation in the favored cosmological model will nevertheless warrant their observations with 30-40-meter class telescopes.

Irrespective of which new strategies are put in place, we can expect that near-field cosmology with dwarf galaxies will need to evolve from an era in which only solidly confirmed dwarf galaxies are used to place constraints on galaxy formation models and/or the properties of dark matter, to an era of statistical samples, with careful considerations for what fraction of the sample of dwarf galaxy candidates could be contaminants. While this will be a paradigm shift and present its own challenges, requiring the assessment of what algorithm can and cannot find, as well as the level of contamination, with detailed mock observations, it presents no fundamental difficulty.

Despite the challenges that will be brought about by deeper surveys, one of their powerful benefits is that they will enable the search of dwarf galaxies beyond the Local Group. Such searches are already being conducted \citep[\eg][]{crnojevic16,davis24,mao24,mutlu-pakdil24,okamoto24} but Euclid, Roman, and LSST will allow for a systematic search of dwarf galaxies around many galaxy hosts within megaparsecs. This will allow for a detailed study of the stochasticity inherent to galaxy formation, as well as the impact of environment, or the properties of the host, on the properties of dwarf galaxies. The intriguing differences observed between the Milky Way and M\,31 dwarf galaxies (see section~\ref{sec:MW-M31_comp}) will likely be better understood once numerous satellite systems are characterized in depth.

We note, however, that even though many new hosts and their surroundings are naturally included in the footprint of upcoming panoptic surveys, it is not the case for M\,31. Our cosmic neighbor is too far north for a large fraction of its halo to be observed by LSST and its location close to both the Galactic and ecliptic plane means that the region out to its virial radius is mainly avoided by the main surveys of both Euclid and CSST. Any opportunity to expand the PAndAS survey deeper and wider (remember that PAndAS only covers the region within half the virial radius of M\,31) should be seized. The M\,31 dwarf galaxy system remains the easiest one to study beyond the Milky Way and we do expect a myriad of satellites are awaiting discovery beyond the edge of PAndAS and/or with a deeper photometric mapping of the PAndAS footprint \citep{dolivadolinsky23}.

Beyond the potential for discovery of dwarf galaxies with the above imaging surveys, new wide-field spectroscopic surveys will soon allow us to understand the chemo-dynamical properties of the known dwarf galaxy population far better. Current and near-future surveys including DESI \citep{cooper23}, WEAVE \citep{jin24}, 4MOST \citep{skuladottir23}, and PFS \citep{takada14} will allow for the measurement of velocities, metallicities and alpha abundances for \emph{thousands} of stars within low mass galaxies around the Milky Way and M\,31. Such a wealth of chemo-dynamic data will allow us to better measure the mass profiles of these systems and place strong, informative constraints on the cusp-core problem and the mass of the dark matter particle. The improved knowledge of the metallicity distribution functions of the dwarf galaxies will also lead to stronger constraints on stellar feedback models through the luminosity-metallicity relation. And, by increasing the number of stars with metallicity measurements in the faintest galaxies, we can better understand if the metallicity plateau at lower luminosities is real, and what causes it.

The next decade therefore presents the opportunity to revolutionise our understanding of the faintest galaxies. With the combination of near-future imaging and spectroscopic surveys, we can (1) deliver a complete census of the LG satellite system down to extremely low masses, (2) extend this to systems beyond the LG, (3) measure precision mass profiles and metallicities in these galaxies and (4) better understand the small-scale issues that have plagued near-field cosmology over the last decade. Over the next decade, this will permit precision constraints on the dark matter particle, and a far better understanding of the baryonic process of galaxy formation. 

\section{Concluding remarks}

In this Chapter, we have reviewed the known satellite galaxy population of the Local Group. This diverse populations of galaxies covers a wide range of luminosities, shapes and morphologies. Thanks to key surveys like the SDSS, PanSTARRS, PAndAS and DES, we have detected tens of low luminosity systems around the Milky Way and M\,31, some of which challenge our ideas of what galaxies really are. Dedicated follow up has allowed us to map out the dark matter components of these systems, leading to constraints on the dark matter model. Their chemical abundances and star formation have illuminated the evolution of the lowest mass galaxies, and have left us with a number of key questions about the nature of galaxy formation. In the near future, we expect many more detections of low-mass galaxies throughout the Local Group and beyond. Such a treasure trove will allow us to put strong constraints on the nature of dark matter, the timing and impact of reionisation, and understand the very limit of galaxy formation. The next decade promises to bring a wealth of breakthroughs in this field. 

\begin{ack}[Acknowledgments]

 MLMC acknowledges support from STFC grants ST/Y002857/1 and ST/Y002865/1.
\end{ack}

\seealso{Small-Scale Challenges to the $\Lambda$CDM Paradigm \citep{bullock17}; The Faintest Dwarf Galaxies \citep{simon19}; Stellar dynamics and dark matter in Local Group dwarf galaxies \citep{battaglia22_2}; Observational constraints on stellar feedback in dwarf galaxies \citep{collins22b}; Baryonic solutions and challenges for cosmological models of dwarf galaxies \citep{sales22};}

\bibliographystyle{Harvard}
\bibliography{reference}

\end{document}